\documentclass[11pt]{article}
\usepackage[a4paper,top=2cm,bottom=2cm,left=2.5cm,right=2.5cm,marginparwidth=1.75cm]{geometry}

\usepackage{amsfonts,amsmath,amssymb,ascmac,bm,tensor}
\usepackage{acronym, cite}

\usepackage{graphicx}
\DeclareGraphicsExtensions{.pdf}
\ifpdf
  \usepackage{color}     %   usepackage without driver option
  \usepackage[bookmarksopen,colorlinks=true,linkcolor=blue,citecolor=blue,urlcolor=magenta]{hyperref}
\else     % For (p)LaTeX + dvipdfmx
\fi

\makeatletter
\let\MYcaption\@makecaption
\makeatother

\usepackage{subcaption}
\makeatletter
\let\@makecaption\MYcaption
\makeatother

\allowdisplaybreaks

\title{Induced gravitational waves probing \\ primordial black hole dark matter with memory burden}

\begin{document}

\begin{titlepage}
\begin{center}

\hfill{\footnotesize KEK-TH-2654}\\
\hfill{\footnotesize KEK-Cosmo-0358}\\
\hfill{\footnotesize KEK-QUP-2024-0021}\\

\vspace{2.6 cm}

{\LARGE Induced Gravitational Waves probing \\ \vskip 1.6 mm Primordial Black Hole Dark Matter with Memory Burden}

\vspace{1.4 cm}

{\large Kazunori Kohri$^{a,b,c,d}$, Takahiro Terada$^{e,f}$, and Tsutomu T.~Yanagida$^{d,g}$}

\vspace{7 mm}

\textit{\small 
${}^{a}$ Division of Science, National Astronomical Observatory of Japan,
Mitaka, Tokyo 181-8588, Japan \\
${}^{b}$ The Graduate University for Advanced Studies (SOKENDAI), Mitaka, Tokyo 181-8588, Japan \\
${}^{c}$ Theory Center, IPNS, and QUP (WPI), KEK, 1-1 Oho, Tsukuba, Ibaraki 305-0801, Japan \\
${}^{d}$ Kavli IPMU (WPI), UTIAS, The University of Tokyo, Kashiwa, Chiba 277-8583, Japan \\
${}^{e}$ Kobayashi-Maskawa Institute for the Origin of Particles and the Universe, Nagoya University, Tokai National Higher Education and Research System, Furo-cho Chikusa-ku, Nagoya 464-8602, Japan \\
${}^{f}$ TRIP Headquarters, RIKEN, Wako 351-0198, Japan \\
${}^{g}$ Tsung-Dao Lee Institute \& School of Physics and Astronomy, Shanghai Jiao Tong University,\\ Pudong New Area, Shanghai 201210, China
}

\vspace{1cm}

\normalsize 

\begin{abstract}
Quantum evaporation of a black hole is conventionally studied semiclassically by assuming self-similarity of the black hole throughout the evaporation process.  However, its validity was recently questioned, and the lifetime of a black hole is conjectured to be much extended by the memory burden effect.   
It gives rise to the possibility that the primordial black holes (PBHs) \emph{lighter} than $10^{10}$ grams are the dark matter in the Universe.  To probe such PBH dark matter, we study gravitational waves (GWs) induced by primordial curvature perturbations that produced the PBHs. 
We find $\Omega_\text{GW}(f_\text{peak})h^2 = 7 \times 10^{-9}$ with the peak frequency $f_\text{peak} = 1\times 10^{3} \, (M_\text{PBH}/(10^{10}\,\mathrm{g}))^{-1/2}\, \mathrm{Hz}$, and the induced GWs associated with the PBH dark matter whose initial mass is greater than about $10^7$ grams can be tested by future observations such as Cosmic Explorer.  
Furthermore, the scenario can be in principle confirmed by detecting another GW signal from the mergers of PBHs, which leads to high-frequency GWs with $f_\text{peak} = 2 \times 10^{27}\, (M_\text{PBH, ini}/(10^{10}\, \mathrm{g}))^{-1}  \, \mathrm{Hz} $. On the other hand, the induced GW signals stronger than expected would contradict the dark matter abundance and exclude the memory burden effect. 
\end{abstract}

\end{center}
\end{titlepage}

\section{Introduction}

Primordial black holes (PBHs) are one of the leading candidates of cold dark matter other than elementary particles. They are classically stable, but they evaporate by the quantum effect emitting Hawking radiation~\cite{Hawking:1974rv, Hawking:1975vcx}.  The lifetime scales as $M_\text{PBH,ini}^3/M_\text{P}^4$ where $M_\text{PBH,ini}$ is the initial mass of the PBH and $M_\text{P}$ is the reduced Planck mass, so the PBHs heavier than about $10^{15}$ gram can survive until today to be the dark matter in the Universe~\cite{Chapline:1975ojl}.  We refer the readers to reviews~\cite{Carr:2020gox, Carr:2020xqk, Green:2020jor, Escriva:2022duf} for the status of PBHs as a dark matter candidate.

The above conventional picture is based on the semiclassical calculation~\cite{Hawking:1975vcx}, in which quantum fields for radiation degrees of freedom are treated on the fixed classical black-hole geometry. Apparently, this makes sense for a sufficiently heavy black hole since the backreaction should be small at each moment.  Though it is a small effect, the black hole gradually loses its mass. Then, one can study the Hawking emission process with a black-hole mass a bit smaller than the initial value.  Supported by the no-hair theorem~\cite{Price:1971fb, Price:1972pw}, one usually assumes the self-similarity of the problem. Namely, one repeats the same argument even when the mass of the black hole significantly changes.  It is usually assumed that the evaporation continues at least until the black hole becomes as light as the Planck mass, so knowledge about quantum gravity is necessary.  This leads to the information loss paradox~\cite{Hawking:1976ra}, which itself is not our main interest in this paper.  Rather, we discuss the cosmological consequences of the potential breakdown of self-similarity and the semiclassical calculation well before reaching the Planck scale.

In series of papers~\cite{Dvali:2011aa, Dvali:2012en, Dvali:2013eja, Dvali:2015aja, Dvali:2017ktv, Dvali:2017nis, Dvali:2018vvx, Dvali:2018tqi, Dvali:2018xpy, Dvali:2020wft, Dvali:2024hsb}, Dvali and his collaborators developed the following picture, in which PBHs are effectively stabilized against Hawking evaporation.  First, a black hole can be understood as a condensate of a large number of soft gravitons~\cite{Dvali:2011aa, Dvali:2012en, Dvali:2013eja}.  A crucial assumption is that these soft modes interact with hard modes in such a way that once a critical number of the soft modes is populated, a large number of hard modes gets gapless. This phenomenon is called assisted gaplessness~\cite{Dvali:2017ktv, Dvali:2017nis, Dvali:2018vvx, Dvali:2018tqi}. The soft mode controls the energy level of the hard modes, so the former is also called the master mode.  The hard modes can encode an exponential amount of information and are also called memory modes, which are mostly responsible for the entropy of the black hole.  A typical microstate of the memory modes contains a large excitation number of gapless modes, whose energy levels depend on the occupation number of the master mode.  This means that the memory modes backreact on the master mode dynamics since the change of the master mode population would cost huge energy~\cite{Dvali:2018xpy, Dvali:2020wft}. Then, the slogan is that a black hole, or more generally, any system with large memory modes is effectively stabilized by the burden of its memory.  The effect gets important at the latest when the black hole loses an $\mathcal{O}(1)$ fraction of its initial mass.\footnote{It may become important as soon as an $\mathcal{O}(1/\sqrt{S_\text{BH}})$ fraction of its initial mass is emitted~\cite{Dvali:2015wca, Michel:2023ydf} where $S_\text{BH}$ is the initial entropy of the black hole.}  Quantitatively, they claimed that this occurs at the latest when the black hole loses half of its initial mass~\cite{Dvali:2020wft}.  That is, an older black hole that emitted more than half of its initial mass is not equivalent to a young black hole with the same mass, violating the self-similarity.

After the memory burden effect becomes important, the precise behavior of the black hole is not clear.  Dvali \textit{et al.}~proposed either of the following possibilities occur~\cite{Dvali:2018xpy, Dvali:2020wft}: (i) the Hawking emission rate is significantly suppressed, or (ii) a new classical instability comes in so that the black hole decays into some lumps and gravitational waves.  The first possibility opens up a new allowed window for the PBH dark matter mass~\cite{Dvali:2018xpy, Dvali:2020wft, Alexandre:2024nuo, Thoss:2024hsr}, so we focus on this possibility.\footnote{
Dark matter can be PBHs in this case but does not have to be so.  Dark matter parameter space in the presence of memory-burden PBHs was studied in Refs.~\cite{Haque:2024eyh, Barman:2024iht}.
}  Quantitatively, Dvali \textit{et al.}~argued that the Hawking emission rate is suppressed by some integer power $n_\text{MB}$ [defined in Eq.~\eqref{MB} below] of its entropy $S_\text{BH}$~\cite{Dvali:2018xpy, Dvali:2020wft, Alexandre:2024nuo}. As the black hole entropy is huge, $S_\text{BH} \gg 1$, the lifetime of a black hole is significantly extended. However, black holes do evaporate with a suppressed Hawking-emission rate, so we are not considering the absolutely stable remnants of black hole evaporation.\footnote{
They are in tension with the covariant entropy bound~\cite{Bousso:1999xy}. For an extended discussion on remnants, see Ref.~\cite{Chen:2014jwq}. Induced gravitational waves to probe black hole remnants were studied in Refs.~\cite{Domenech:2023mqk, Franciolini:2023osw}. 
} 

%====================================================================
\begin{figure}[tbh]
    \centering
    \includegraphics[width=0.49\textwidth]{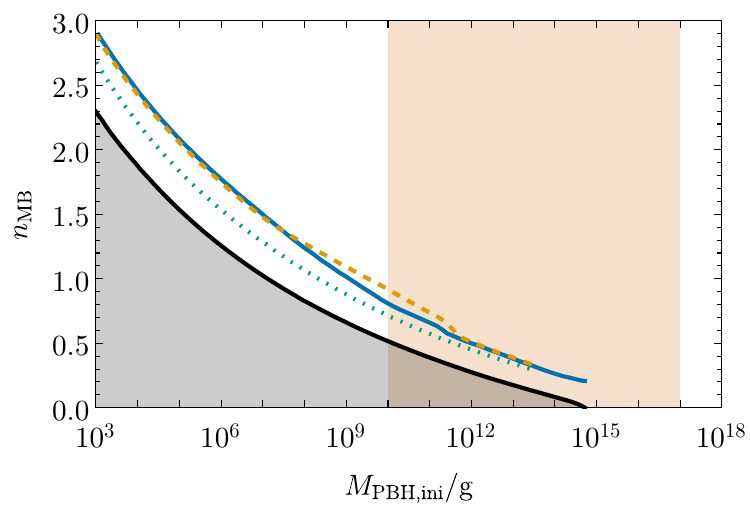}
    \includegraphics[width=0.49\textwidth]{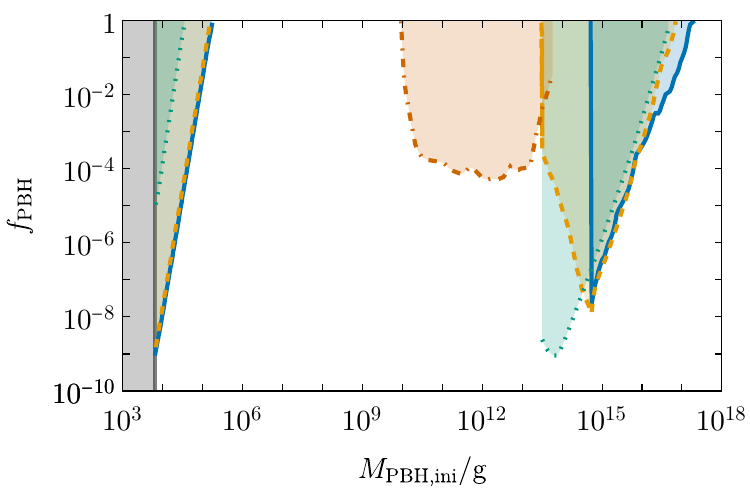}
    \caption{Constraints on the PBH parameter space with the memory burden effect adopted from Ref.~\cite{Thoss:2024hsr}. Left: the blue solid, orange dashed, and green dotted lines are the lower bounds on the memory-burden parameter $n_\text{MB}$ from the galactic gamma-ray background, extragalactic gamma-ray background, and cosmic microwave background (CMB), respectively. In the black shaded region, PBHs have evaporated by the present time even with the memory burden effect. The range $10^{10}\,\mathrm{g} \leq M_\text{PBH,ini} \leq 10^{17}\,\mathrm{g}$ is shaded in vermilion to approximately show the constraints from the BBN. Right: the upper bound on the PBH abundance $f_\text{PBH}$ in the case of $n_\text{MB} = 2$. The blue solid, orange dashed, green dotted, and vermilion dot-dashed lines are the constraints from the galactic gamma-ray background, extragalactic gamma-ray background, CMB, and BBN, respectively.  To the left of the black vertical line, the PBHs have evaporated by the present time.  See Ref.~\cite{Thoss:2024hsr} for more details.
 }
    \label{fig:constraints}
\end{figure}
%====================================================================

Cosmological constraints on PBHs beyond the semiclassical regime, i.e., with the memory burden effect were studied in Refs.~\cite{Alexandre:2024nuo, Thoss:2024hsr}.  In particular, Ref.~\cite{Thoss:2024hsr} showed the upper bound on the initial PBH mass of $M_\text{PBH,ini} \lesssim 10^{10} \, \mathrm{g}$ from big bang nucleosynthesis (BBN) while the lower bound from galactic and extragalactic gamma rays \cite{Carr:2009jm,Carr:2020gox} depends on the suppression power $n_\text{MB}$ of the memory burden effect (see the left panel of Fig.~\ref{fig:constraints}). 
For $n_\text{MB} = 1$, the allowed region is almost closed.  For $n_\text{MB} = 2$, the allowed mass is $10^5 \, \mathrm{g} \lesssim M_\text{PBH,ini} \lesssim 10^{10}\, \mathrm{g}$ (see the right as well as left panel of Fig.~\ref{fig:constraints}).  While $n_\text{MB} > 2$ allows lighter PBHs and this extension is straightforward, the associated induced gravitational waves (GWs) (introduced in the next paragraph) have too high frequencies to be detected in the near future.  Assuming the post-inflationary PBH formation mechanism in the radiation-dominated era~\cite{Hawking:1971ei, Carr:1974nx}, the lower bound on the PBH mass is $M_\text{PBH,ini} \gtrsim 0.5 \, \text{g}$, which comes from the upper bound on the tensor-to-scalar ratio $r \lesssim 0.04$~\cite{BICEP:2021xfz, Tristram:2021tvh, Paoletti:2022anb} with the measured amplitude of the primordial curvature perturbations $A_\text{s} = 2.1 \times 10^{-9}$~\cite{Planck:2018vyg, Planck:2018jri}.  Therefore, we consider the mass range $0.5 \, \mathrm{g} \lesssim M_\text{PBH,ini} \lesssim 10^{10}\, \mathrm{g}$ but focus on $10^5 \, \mathrm{g} \lesssim M_\text{PBH,ini} \lesssim 10^{10}\, \mathrm{g}$.

As pointed out in Refs.~\cite{Saito:2008jc, Saito:2009jt}, the PBH scenario may well be associated with GWs induced by the primordial curvature perturbations. This is because, in a popular PBH production mechanism, it is produced by the gravitational collapse of a cosmological patch with extremely enhanced curvature perturbations~\cite{Hawking:1971ei, Carr:1974nx}. 
The latter necessarily induces a significant intensity of gravitational waves beyond the linear order of the cosmological perturbation theory~\cite{1967PThPh..37..831T, Matarrese:1992rp, Matarrese:1993zf, Matarrese:1997ay, Ananda:2006af, Baumann:2007zm}. It is useful to note that the initial PBH mass and the typical frequency of the induced GWs are in one-to-one correspondence. In the conventional scenario without the memory burden effect, the asteroid-mass PBHs can constitute 100\% abundance of the dark matter, whose corresponding induced GWs can be probed by a space-based GW interferometer~\cite{Saito:2008jc, Saito:2009jt} such as Laser Interferometer Space Antenna (LISA)~\cite{LISA:2017pwj, Baker:2019nia}.

In this paper, we study the induced GWs associated with the PBH dark matter in the presence of the memory burden effect.\footnote{GWs from Hawking radiation of PBHs in the presence of the memory burden effect were studied in Ref.~\cite{Barman:2024ufm}, which appeared on arXiv simultaneously with this paper.}  Our focus is the simple scenario in which the induced GWs are produced around the time of PBH formation in the radiation-dominated era.\footnote{
Recently, another induced-GW signal related to the \emph{non}-dark-matter PBHs was studied in Refs.~\cite{Balaji:2024hpu, Barman:2024iht, Bhaumik:2024qzd}. 
In their scenario, PBHs dominate the Universe (see also Ref.~\cite{Domenech:2021and} for the GWs induced by PBH isocurvature perturbations) and then evaporate to reheat the Universe. If the distributions of the mass and spin of PBHs are negligible, the PBHs evaporate almost simultaneously leading to a rapid change of the equation of state of the Universe. In such an occasion, the induced GWs are significantly enhanced by the poltergeist mechanism~\cite{Inomata:2019ivs} (named in Ref.~\cite{Inomata:2020lmk}), where the perturbations (of the PBH number density~\cite{Inomata:2020lmk}) were assumed to be adiabatic. It was generalized to the Poissonian case~\cite{Papanikolaou:2020qtd} in Ref.~\cite{Domenech:2020ssp}. The Poissonian contribution was considered in Refs.~\cite{Balaji:2024hpu, Barman:2024iht}, while both contributions were considered in Ref.~\cite{Bhaumik:2024qzd}.  See also Refs.~\cite{Inomata:2019zqy, Inomata:2020lmk, Papanikolaou:2022chm, Harigaya:2023mhl, Pearce:2023kxp} for the importance of the finite transition timescale, which makes the enhancement of the induced GWs milder.  Note also that the mass distribution of the memory-burden PBHs was discussed in Ref.~\cite{Dvali:2024hsb}, which suppresses the induced GWs from the poltergeist mechanism~\cite{Inomata:2020lmk, Papanikolaou:2022chm}.
}
In Sec.~\ref{sec:PBH}, we review PBHs discussing their formation, evaporation, and abundance in the presence of the memory burden effect. We introduce example power spectra of the primordial curvature perturbations.  Sec.~\ref{sec:GW} is the main section of this paper, where we discuss the induced-GW signals of the memory-burdened PBH dark matter scenario.  We discuss the observational prospects of the induced GWs.  In Sec.~\ref{sec:tests}, we discuss some possibilities to cross-check or exclude the memory-burdened PBH dark matter scenario. For example, we argue that the memory burden effect can be excluded if the induced GWs associated with the light PBHs are too strong. We also discuss another GW signal, which is from mergers of PBHs, to confirm the memory burden effect although the frequency of the GWs is extremely high. We summarize the paper and conclude in Sec.~\ref{sec:conclusion}. We use the natural unit $c = \hbar = k_\text{B} = 8\pi G = 1$ unless explicitly denoted.

\section{Primordial black hole abundance with memory burden\label{sec:PBH}}
In this section, we summarize our method of calculation of the PBH abundance and mention some differences from the conventional case without the memory burden effect.

Although the PBH abundance is sensitive to the details of the calculation method,\footnote{
See Refs.~\cite{Ando:2018qdb, Young:2019osy, Yoo:2020dkz, DeLuca:2023tun, Franciolini:2023wun, Ianniccari:2024bkh} for recent discussions on (uncertainties of) the PBH abundance.  For more general reviews of PBHs, see Refs.~\cite{Sasaki:2018dmp, Carr:2020gox, Carr:2020xqk, Escriva:2022duf}.
} the overall difference can be absorbed by the amplitude of the curvature perturbations.  Since the PBH abundance is extremely sensitive to it while the intensity of the induced GWs, which is the main topic of this paper, is only quadratically sensitive, our results do not depend on the details of the calculation formalism of the PBH abundance. Therefore, we take a simple method; we basically follow the method in Refs.~\cite{Inomata:2017okj, Ando:2018qdb, NANOGrav:2023hvm, Inomata:2023zup}.\footnote{
The differences are as follows.  We use $\delta_\text{c} = 0.42$ rather than $0.45$.  This means that PBHs are more easily produced. To adjust the PBH abundance, a smaller amplitude of the curvature perturbations is required.  This leads to a conservative estimate for the induced GWs. We also use $\gamma = (1/\sqrt{3})^3$ rather than $0.2$ following Ref.~\cite{Carr:1975qj}.  
}

We consider PBH formation in a radiation-dominated epoch. 
The (initial) mass of a PBH created in a Hubble patch with enhanced curvature perturbations is given by a fraction $\gamma$ of the Hubble horizon mass
\begin{align}
    M_\text{PBH,ini} = \frac{4\pi \gamma}{3} H^{-3},
\end{align}
where $H$ is the Hubble parameter, which is related via $H = k/a$ to the wavenumber $k$ of the fluctuation with $a$ being the scale factor of the Friedmann–Lema\^{i}tre–Robertson–Walker metric. We take $\gamma = (1/\sqrt{3})^3$~\cite{Carr:1975qj}. 

Carr's formula~\cite{Carr:1975qj} (cf.~the Press-Schechter formalism~\cite{Press:1973iz}) for the PBH formation probability $\beta$ is
\begin{align}
    \beta(M) = \frac{1}{2} \mathrm{Erfc} \left( \frac{\delta_\text{c}}{\sqrt{2 \sigma^2(k)}}\right),
\end{align}
where Erfc denotes the complementary error function, $\delta_\text{c}$ is the critical overdensity to form a PBH, and $\sigma^2(k)$ is the variance of the coarse-grained overdensity:
\begin{align}
    \sigma^2(k) = \frac{16}{81} \int_{-\infty}^\infty \mathrm{d} \ln q \left( \frac{q}{k} \right)^4 W\left( \frac{q}{k}\right)^2 T\left( \frac{q}{k} \right)^2  \mathcal{P}_\zeta(q), \label{sigma^2}
\end{align}
where $W(z)$ is the window function, for which we take $W(z) = \exp (-z^2 / 2)$, $T(z) = 9\sqrt{3} (\sin (z/\sqrt{3})-z/\sqrt{3} \cos (z/\sqrt{3}) )/z^3$ is the transfer function, and $\mathcal{P}_\zeta(k)$ is the dimensionless power spectrum of the primordial curvature perturbations. 

In this paper, we consider two types of $\mathcal{P}_\zeta (k)$ for illustrative purposes. One is the delta function and the other is Gaussian in terms of $\ln k$. 
\begin{align}
    \mathcal{P}_\zeta (k) = \begin{cases}
        A_\zeta  \delta \left(\ln  \frac{k}{k_0} \right)  \\
        \frac{A_\zeta}{\sqrt{2\pi \Delta^2}} \exp \left( - \frac{\left(\ln \frac{k}{k_0} \right)^2 }{2 \Delta^2} \right)
    \end{cases}
\end{align}
where $k_0$ is a reference scale, $A_\zeta$ is the amplitude of the curvature perturbations, and $\Delta$ is the width of the Gaussian. 
Note that the delta-function case is the $\Delta \to 0$ limit of the lognormal case.
An advantage of the delta function $\mathcal{P}_\zeta (k)$ is that the resulting PBH mass function is as narrow as possible so that the comparison to observational constraints (usually derived by assuming the monochromatic PBH mass function) is relatively straightforward. A disadvantage of the delta-function case is that model building to realize such a narrow limit of $\mathcal{P}_\zeta$ is nontrivial.  On the other hand, an advantage of the lognormal $\mathcal{P}_\zeta (k)$ is that it can approximate realistic smooth spectra.  A disadvantage, however, is that the resulting PBH mass function is broader, so the comparison to PBH constraints becomes nontrivial.  Thus, the two options of $\mathcal{P}_\zeta (k)$ are complementary.

To discuss the PBH abundance at present, we need to discuss Hawking radiation and the memory burden effect.
The mass loss rate by the Hawking radiation is given by (see Refs.~\cite{Hooper:2019gtx, MacGibbon:1990zk, MacGibbon:1991tj} for the case without the memory burden effect) 
\begin{align}
    \dot{M}_\text{PBH} =&  - \frac{\pi \mathcal{G} g_{*\text{H}}}{480 M_\text{PBH}^2} \times \begin{cases}
      1  & (M_\text{PBH} \geq q M_\text{PBH,ini}) \\
       S_\text{BH}^{-n_\text{MB}} & (M_\text{PBH} < q M_\text{PBH,ini})
    \end{cases}
    \label{MB}
\end{align}
where $\mathcal{G} \simeq 3.8$ is the gray-body factor, $g_{*\text{H}}(\simeq 108$ for $M_\text{PBH}\ll 10^{11} \text{g}$ in the Standard Model~\cite{Hooper:2019gtx} at least in the conventional case without memory burden) is the effective number of degrees of freedom emitted by Hawking radiation, and $1/2 \leq q \leq 1$ parametrizes the critical mass at which the memory burden effect becomes relevant.  The Bekenstein-Hawking entropy of the black hole is given by $S_\text{BH}(M_\text{PBH}) = 2 \pi A = M_\text{PBH}^2 / 2$~\cite{Bekenstein:1972tm, Hawking:1975vcx} in the reduced Planck unit with $A = M_\text{PBH}^2 / (4 \pi)$ being the surface area.  There are some caveats in the above expression. First, the detailed form of the emission rate in the memory-burden regime is unknown. Beyond the semiclassical regime, it is not clear if we can rely on notions of the black-hole thermodynamics~\cite{Bardeen:1973gs} such as Hawking temperature and entropy.  Second, the above evaporation rate is discontinuous at $M_\text{PBH} = q M_\text{PBH,ini}$.  Realistically, the evaporation rate will smoothly transform from the semiclassical regime to the memory burden regime because the magnitude of the backreaction from the memory modes to the master mode is continuous.  Third, there are several proposals~\cite{Alexandre:2024nuo, Thoss:2024hsr, Balaji:2024hpu} for the detailed form of the evaporation rate in the memory burden regime. For example, some authors replace $M_\text{PBH}$ and/or $S_\text{BH}(M_\text{PBH})$ with $q M_\text{PBH,ini}$ and/or $S_\text{BH}(q M_\text{PBH,ini})$. These differences lead to only negligible effects on observables at present at least in the parameter range of our main interest $10^5 \leq M_\text{PBH,ini}/\mathrm{g} \leq 10^{10}$. For $n_\text{MB} \geq 2$, it is a good approximation to set $M_\text{PBH}= q M_\text{PBH,ini}$ at present since the Hawking emission rate in the semiclassical regime for these masses is so effective while the evaporation rate is significantly suppressed in the memory burden regime.  

The present abundance of PBHs is that in the conventional case times the fraction $q$ for the parameter range of our main interest
\begin{align}
    f_\text{PBH} \equiv \frac{\rho_\text{PBH}}{\rho_\text{CDM}} = q \int_{-\infty}^\infty \mathrm{d}\ln M \frac{g_{*}(T(M))}{g_{*,0}}\frac{g_{*s,0}}{g_{*s}(T(M))} \frac{T(M)}{T_\text{eq}} \gamma \beta (M) \frac{\Omega_\text{m}}{\Omega_\text{CDM}} ,
\end{align}
where $g_*(T)$ and $g_{*s}(T)$ are the effective number of relativistic degrees of freedom for energy density and entropy density, respectively, $T(M)$ is the temperature at the formation of PBH with its (initial) mass $M$, $T_\text{eq}$ is the temperature at the matter-radiation equality, and $\Omega_X = \rho_X / \rho_\text{total}$ is the energy-density fraction of component X (m: matter, CDM: cold dark matter, r: radiation, etc.). In the following, we fix $q = 1/2$ for definiteness. We use the numerical fitting functions in Ref.~\cite{Saikawa:2018rcs} for $g_*(T)$ and $g_{*s}(T)$.

\section{Induced gravitational waves\label{sec:GW}}
The memory burden effect changes the relation between the initial PBH mass $M_\text{PBH,ini}$ and its present value $M_\text{PBH}$, but the relation between the PBH formation and the induced GWs is unaffected. The enhanced primordial curvature perturbations responsible for the PBH formation also produce GWs secondarily via interactions in General Relativity. Although the leading-order contribution to the induced GWs is the second order, it can dominate over the first-order GWs produced during inflation directly from tensor-field vacuum fluctuation without violating perturbativity.  This is because the source of the first-order (primordial) GWs and the second-order (induced) GWs are different. 

The spectrum of the scalar-induced GWs is obtained as~\cite{Ananda:2006af, Baumann:2007zm, Espinosa:2018eve, Kohri:2018awv} 
\begin{align}
    \Omega_\text{GW}(\eta_0, f) h^2 = & \frac{g_{*}(T(f))}{g_{*,0}} \left( \frac{g_{*s,0}}{g_{*s}(T(f))} \right)^{4/3}  \Omega_\text{r}h^2  \Omega_\text{GW} (\eta_\text{c}, f), \\
    \Omega_\text{GW}(\eta_\text{c}, f) = & \frac{1}{6} \int_0^\infty \mathrm{d}t \int_0^1 \mathrm{d} s  \left( \frac{t(2+t)(s^2-1)}{(1-s+t)(1+s+t)} \right)^2 \lim_{\eta\to \infty}\overline{(k\eta)^2 I(t,s, k\eta)^2} \mathcal{P}_\zeta (u k) \mathcal{P}_\zeta (v k),  
\end{align}
where $\eta$ is the conformal time, $h = H_0 / (100 \, \mathrm{km}/\mathrm{s}/\mathrm{Mpc})$ is the reduced Hubble parameter, $k = 2 \pi f$ is the wavenumber of the GWs, 
$\Omega_\text{GW}(\eta_\text{c})$ is the value of $\Omega_\text{GW}(\eta)$ that reached a constant value during the radiation-dominated epoch, $u = (t+s+1)/2$, $v = (t-s+1)/2$, and the oscillation average (denoted by the overline) of the kernel function $\lim_{\eta\to \infty}\overline{(k\eta)^2 I(t,s,k \eta)^2}$ is~\cite{Kohri:2018awv, Espinosa:2018eve}
\begin{align}
    &\lim_{\eta\to \infty}\overline{(k\eta)^2 I(t,s, k\eta)^2} = \frac{288 (-5 + s^2 + t(2+t))^2}{(1-s+t)^6 (1 + s +t )^6} \left( \frac{\pi^2}{4} \left(-5 + s^2 + t(2+t)\right)^2 \Theta (t - (\sqrt{3}-1)) \right.  \nonumber \\
     & \qquad \qquad \left. + \left(- (t-s+1)(t+s+1) + \frac{1}{2} \left( -5 + s^2 + t(2+t)\right) \ln \left| \frac{t(2+t) - 2}{3-s^2} \right| \right)^2 \right), 
\end{align}
where $\Theta$ is the Heaviside step function. 

%====================================================================
\begin{figure}[tbh]
    \centering
    \includegraphics[width=0.49\textwidth]{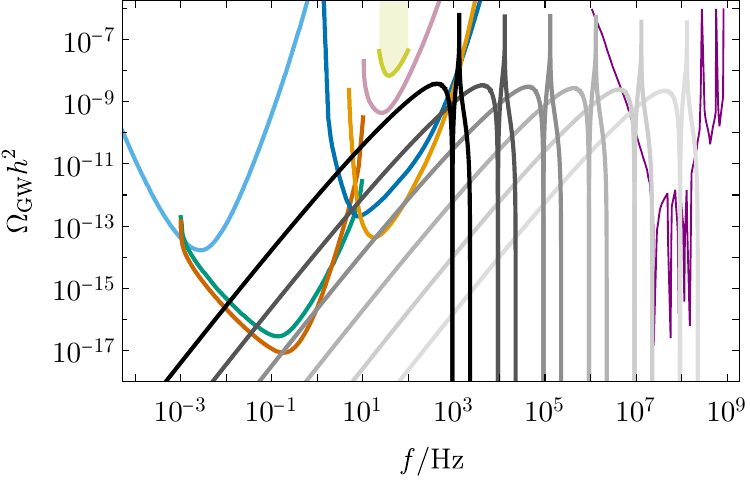}
    \includegraphics[width=0.49\textwidth]{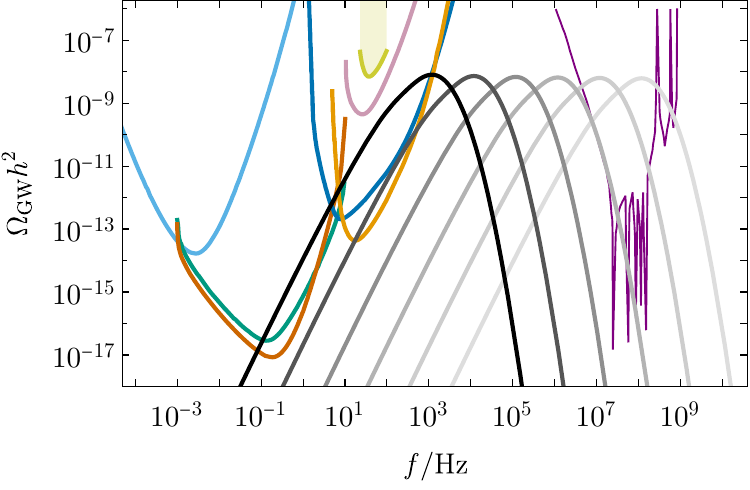}
    \caption{
    The spectra of the induced GWs associated with the PBH dark matter are shown in monochrome colors. From left (black) to right (light gray), PBHs with $2 M_\text{PBH}/\mathrm{g} =  M_\text{PBH,ini}/\mathrm{g} = 10^{10}$, $10^8$, $10^6$, $10^4$, $10^2$, and $10^0$ are assumed to comprise the totality of dark matter with the help of the memory burden effect.  The left and right panels show the cases with $\mathcal{P}_\zeta$ being the delta function and the lognormal function with $\Delta = 1$, respectively. The colored lines in the region $f<10^4 \, \mathrm{Hz}$ denote the power-law integrated sensitivity curves for 1-year observations adopted from Ref.~\cite{Schmitz:2020syl}. The observations include LISA (sky blue), DECIGO (bluish green), BBO (vermilion), ET (blue), CE (orange), HLVK (design sensitivity) (reddish purple), and HLV O3 (yellow shaded). In addition, the sensitivity of future resonant cavities~\cite{Berlin:2021txa, Herman:2022fau}, assuming the detectability of the strain of $10^{-14}\,\mathrm{W}$, is adopted from Ref.~\cite{Jiang:2024akb} and shown by the thin purple line.
 }
    \label{fig:GW_spectra}
\end{figure}
%====================================================================

The spectra of the induced GWs are plotted in Fig.~\ref{fig:GW_spectra} with various observational sensitivity curves. The left panel shows the case of the delta-function $\mathcal{P}_\zeta (k)$, and the right panel shows the case of the lognormal $\mathcal{P}_\zeta (k)$.  Since a small-scale mode re-enters the Hubble horizon earlier, the induced GWs associated with a smaller PBH have higher frequencies.  For simplicity, we set the present PBH mass to be the half of its initial value $M_\text{PBH} = M_\text{PBH,ini}/2$ for all choices of $M_\text{PBH,ini}$ in the figure.  

The largest difference between the delta-function case and the lognormal case is the power-law index of $\Omega_\text{GW}(f)$ on the infrared (IR) tail.  The power-law index is $2$ and $3$ for the delta-function case and the lognormal case, respectively, up to a logarithmic correction~\cite{Yuan:2019wwo}.  The latter value $3$ is the universal result based on causality and statistics with the assumption that the GWs are emitted in a finite duration in the radiation-dominated epoch~\cite{Cai:2019cdl}.  For details, see also Ref.~\cite{Pi:2020otn}. Because of the lightness of the memory-burdened PBH dark matter, the mainly observable part of the GW spectrum is the IR tail, whereas the peak height of the GW spectrum is approximately the same for both cases once we fit the PBH dark matter abundance ($f_\text{PBH}=1$). Therefore, the induced GWs from the delta-function $\mathcal{P}_\zeta (k)$ can be more easily tested. 

The colored lines in the region $f < 10^4 \,\mathrm{Hz}$ in Fig.~\ref{fig:GW_spectra} are the power-law integrated sensitivity curves of the following GW observations assuming 1-year observations. The sky blue, bluish green, and vermilion lines correspond to LISA~\cite{LISA:2017pwj, Baker:2019nia}, DECi-hertz Interferometer Gravitational wave Observatory (DECIGO)~\cite{Seto:2001qf, Yagi:2011wg, Isoyama:2018rjb, Kawamura:2020pcg}, and Big Bang Observer (BBO)~\cite{Crowder:2005nr, Corbin:2005ny, Harry:2006fi}.  These are space-based interferometers. The shaded yellow region is the Laser Interferometer Gravitational-Wave Observatory (LIGO)/Virgo (HLV) O3 constraint on SGWB~\cite{KAGRA:2021kbb}. This is an existing constraint whereas the other lines are sensitivity curves for future observations taken from Ref.~\cite{Schmitz:2020syl}. The pink line corresponds to a combination of advanced LIGO~\cite{Harry:2010zz, LIGOScientific:2014pky}, advanced Virgo~\cite{VIRGO:2014yos}, and Kamioka Gravitational wave detector (KAGRA)~\cite{Somiya:2011np, Aso:2013eba, KAGRA:2018plz, KAGRA:2019htd, Michimura:2019cvl} (HLVK). These are ground-based interferometers. The next generation ground-based interferometers include Einstein Telescope (ET)~\cite{Punturo:2010zz, Hild:2010id, Sathyaprakash:2012jk, Maggiore:2019uih} (the blue line) and  Cosmic Explorer (CE)~\cite{LIGOScientific:2016wof, Reitze:2019iox} (the orange line). For the reference of  high-frequency GW observations, we also plot the sensitivity~\cite{Herman:2022fau, Jiang:2024akb} of futuristic resonant cavities~\cite{Berlin:2021txa, Herman:2022fau} by the thin purple line in the region $f > 10^5 \, \mathrm{Hz}$, assuming the detectability of the strain of $10^{-14}\,\mathrm{W}$.

%====================================================================
\begin{figure}[tbh]
    \centering
    \includegraphics[width=0.49\textwidth]{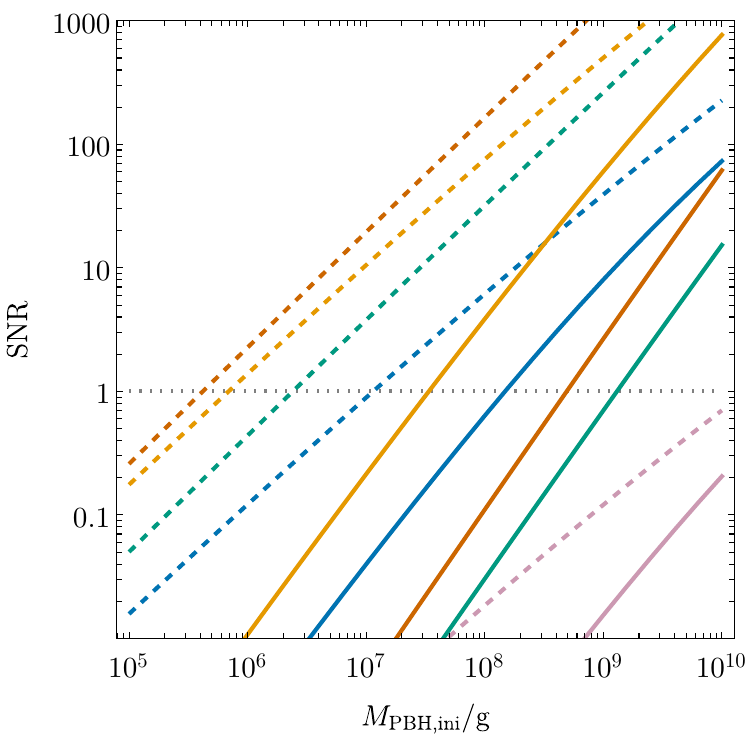}
    \includegraphics[width=0.49\textwidth]{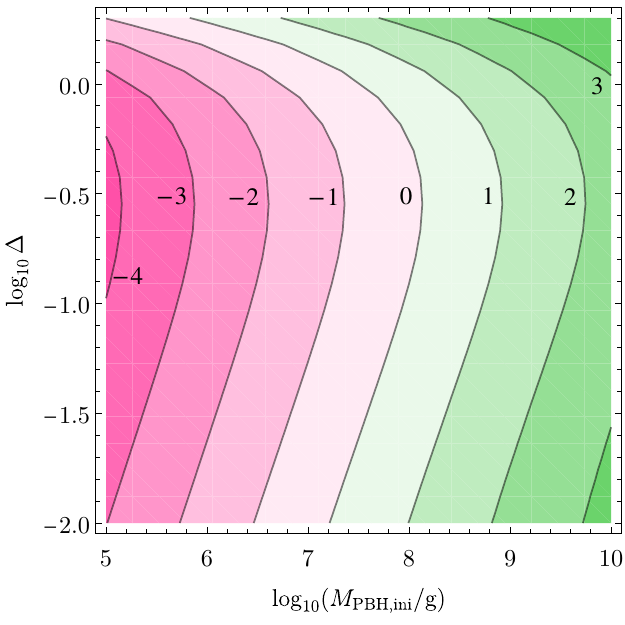}
    \caption{
    The signal-to-noise ratio (SNR) for 1-year observations. Left: the SNR as a function of the initial PBH mass. The solid and dashed lines correspond to the cases with $\mathcal{P}_\zeta( \ln k)$ being the Gaussian with $\Delta = 1$ or the delta function, respectively. The orange, blue, vermilion, bluish-green, and reddish-purple lines correspond to CE, ET, BBO, DECIGO, and HLVK, respectively. Right: the contour plot of $\log_{10} \text{SNR}$ for CE (1 year) on the plane of $\log_{10}(M_\text{PBH,ini}/\text{g})$ and $\log_{10}\Delta$. The contour lines denote $-4$, $-3$, \dots, $3$ from left to right; the border between the pinkish and greenish regions corresponds to $\text{SNR} = 1$. See Appendix~\ref{sec:SNR} for similar plots for ET, BBO, DECIGO, and HLVK.
 }
    \label{fig:SNR}
\end{figure}
%====================================================================

Let us more quantitatively discuss the observational prospect. The IR tail of the GW spectrum approximately follows the power law, so the power-law integrated sensitivity curves are useful guides.  However, the signals do not completely follow the power law.  To be more precise, it is better to compute the signal-to-noise ratio (SNR) without assuming the power law. It is defined as follows~\cite{Maggiore:1999vm, Allen:1996vm, Allen:1997ad}
\begin{align}
    \text{SNR} =& \sqrt{n_\text{det} t_\text{obs} \int_{f_\text{min}}^{f_\text{max}} \mathrm{d}f \left( \frac{\Omega_\text{signal}(f)}{\Omega_\text{noise}(f)} \right)^2 },
\end{align}
where $n_\text{det}$ is the number of detectors in taking correlations, $t_\text{obs}$ is the observational time, $f_\text{max}$ and $f_\text{min}$ are the maximum and minimum frequencies for each observation, and  $\Omega_\text{signal}(f)$ and $\Omega_\text{noise}(f)$ are the signal and noise spectra, respectively.\footnote{
We use the data of the noise spectra available at \url{https://zenodo.org/records/3689582} associated with Ref.~\cite{Schmitz:2020syl}. $f_\text{min}$ and $f_\text{max}$ are identified with the minimum and maximum frequencies in each dataset.
} We set $n_\text{det} = 2, 2, 2, 1$, and $4$ for BBO, DECIGO, CE, ET, and HLVK, respectively. 

The SNR in our setup is shown in Fig.~\ref{fig:SNR}. The left panel shows the SNR for each observation.  The solid lines are for the lognormal $\mathcal{P}_\zeta (k)$ with $\Delta = 1$ whereas the dashed lines are for the delta-function $\mathcal{P}_\zeta (k)$. The orange, blue, vermilion, bluish-green, and reddish-purple lines correspond to CE, ET, BBO, DECIGO, and HLVK, respectively. As we discussed above, the signal can be more easily detected in the delta-function case because of the smaller power-law index of the IR tail. Since the space-based interferometers (DECIGO and BBO) sit in the deep IR, the increase of SNR of these observations is more significant from the lognormal case (solid lines) to the delta-function case (dashed lines).

The right panel of Fig.~\ref{fig:SNR} is the contour plot of $\log_{10}\text{SNR}$ for CE on the plane spanned by $\log_{10} (M_\text{PBH,ini}/\text{g})$ and $\log_{10}\Delta$.  The corresponding contour plots for the other GW observations look qualitatively the same (see Appendix~\ref{sec:SNR}). The figure again shows that sufficiently heavy memory-burdened PBH dark matter can be detected because of their relatively low-frequency induced GWs. Now, let us focus on the dependence on $\Delta$. Since $\Delta$ controls the spectral width in terms of $\ln k$, the dependence of the SNR on $\Delta$ is significant for $\Delta \gtrsim 1$; the broader the spectra, the easier the observation is.  For $\Delta \lesssim 1$, however, decreasing $\Delta$ does not narrow the GW spectrum because of the existence of the IR tail in the induced GW spectrum.  Moreover, the power-law index of the IR tail approaches $2$ in the limit $\Delta \to 0$, so the sensitivity becomes better again toward the bottom of the figure.  This explains the non-monotonic dependence of the SNR on $\Delta$.

\section{Tests of the scenario\label{sec:tests}}

Detection of the induced GWs with their intensity at the expected level (see Fig.~\ref{fig:GW_spectra}) gives us affirmative information for the production of PBHs in the early Universe. However, it does not give us any hints on whether the produced PBHs have evaporated away or constitute dark matter at present. In this section, we discuss ways to confirm or exclude the scenario of the memory-burdened PBH dark matter. 

\subsection{Way to exclude the memory burden effect}

If the induced GWs are observed with their intensity greater than the expected level, it means that the abundance of PBHs at their formation is greater than what is needed to explain the dark matter abundance. In the memory burden scenario, they lead to the overproduction of PBH dark matter. 
Even a signal a few times or dozens of percent greater than the expected level implies orders of magnitude too many PBHs. Therefore, if the induced GWs associated with light PBHs stronger than expected are observed, one can exclude the memory burden effect.

There is a caveat on this program. If one observes only the IR tail of the GW spectrum, the information of the amplitude and of the peak frequency, or equivalently, the PBH abundance and PBH mass, are degenerate because a smaller amplitude with a lower peak frequency has a similar signal with a larger amplitude with a higher peak frequency.  To exclude the memory burden effect, one first needs to measure the GW spectrum in a sufficiently broad range so that the degeneracy is resolved. 

Suppose that one has excluded the memory burden effect by observing the induced GWs stronger than expected.  In this case, if the PBHs were so abundant that they became the dominant component of the Universe, their evaporation produce the hot big-bang Universe associated with gravitons, which contribute to the effective number of species of neutrinos $N_\text{eff}$.  The excess can be detected by future observations such as CMB-S4~\cite{CMB-S4:2016ple} if the PBHs acquire substantial spin, e.g., by merger processes~\cite{Hooper:2020evu, Cheek:2022mmy}.

On the other hand, a way to avoid  exclusion of the memory burden effect is to dilute the overproduced memory-burdened PBH dark matter by entropy production due, e.g., to heavy particle decay (see, e.g., Refs.~\cite{Coughlan:1983ci, Banks:1993en, deCarlos:1993wie, Kawasaki:1999na, Buchmuller:2006tt}) or to thermal inflation~\cite{Lyth:1995ka}, though this also dilutes the baryon asymmetry of the Universe if the generation of the baryon asymmetry took place earlier.  In addition, the (incomplete) evaporation of the PBHs produces entropy and dilutes the baryon asymmetry further.  The PBHs initially heavier than about $10^9\,\mathrm{g}$ are tightly constrained by the BBN.  If one assumes that the baryon asymmetry of the Universe was produced sufficiently early (before PBH evaporation), the observed baryon-to-photon ratio puts an upper bound on the initial PBH abundance $\gamma \beta = (\rho_\text{PBH}/\rho_\text{total})|_\text{ini}$: $\beta' \lesssim 10^{-5} \left(\frac{10^9 \, \mathrm{g}}{M_\text{PBH,ini}} \right)$ for $M_\text{PBH,ini} < 10^9\, \mathrm{g}$, where $\beta' \equiv \gamma^{3/2} \left( \frac{106.75}{g_{*,\text{ini}}}\right)^{1/4} \left(\frac{0.67}{h}\right)^2 \beta$  (see Ref.~\cite{Carr:2020gox} and references therein).\footnote{
The definition of $\beta$ in this paper is different from that in Ref.~\cite{Carr:2020gox}: $\gamma \beta_\text{here} = \beta_\text{there}$, where $\beta_\text{here}$ is  $\beta$ in this paper (the formation probability) and $\beta_\text{there}$ is $\beta$ defined in Ref.~\cite{Carr:2020gox} (the initial abundance). 
} Note that this bound can be evaded if the baryon asymmetry of the Universe is produced after the PBH evaporation and before the BBN. 

In this way, one can constrain either the memory burden effect or the generation mechanism of the baryon asymmetry of the Universe if the too-strong induced GW signal is observed.

\subsection{Way to confirm the memory burden effect}
Because the PBHs lighter than $10^{15}\,\mathrm{g}$ evaporate away in the conventional scenario, the phenomenology of dark matter in the mass range $M_\text{P} \leq M \lesssim 10^{15}$ has not been studied as thoroughly as the heavier case $M \gtrsim 10^{15}\,\mathrm{g}$.  Note also that the memory-burdened PBH in the memory burden phase typically has a mass of the order of its initial mass in contrast to the Planck-mass relics. Thus, the memory-burdened PBH dark matter and its properties such as mass and number density are in between those of particle dark matter and the conventional PBH dark matter without the memory burden effect. There may be unexplored rich phenomenology, and its thorough exploration is left for future work.  

Let us discuss the GWs originating from mergers of the memory-burdened PBH dark matter.  
This has been well studied in the context of explaining the LIGO/Virgo GW signals by PBH binary mergers. PBHs can form a binary in the early Universe~\cite{Nakamura:1997sm, Ioka:1998nz, Sasaki:2016jop, Sasaki:2018dmp} as well as in the late Universe~\cite{Bird:2016dcv, Clesse:2016vqa}.  For $30\, M_\odot$ PBHs, the former mechanism is more effective~\cite{Sasaki:2016jop, Sasaki:2018dmp}. The merger rate per comoving volume approximately scales as $M_\text{PBH}^{-32/37}$ and $M_\text{PBH}^0$ for the early and late formation mechanisms, respectively, so the early-time binary formation mechanism is dominant also for PBHs lighter than $30 \, M_\odot$. The spectrum of the stochastic GW background from the superposition of mergers of light PBH binaries was discussed in the literature~\cite{Wang:2016ana, Wang:2019kaf, Inomata:2020lmk, Kohri:2020qqd, Inomata:2023zup}.\footnote{
The same calculation applies to the supermassive black hole binaries if the environmental effect is negligible.  If the evidence~\cite{NANOGrav:2023gor, EPTA:2023fyk, Reardon:2023gzh, Xu:2023wog} of the GWs around nanohertz found by the worldwide pulsar timing array collaborations originates from the mergers of the supermassive black hole binaries, the environmental effect must be significant. See Refs.~\cite{NANOGrav:2023hfp, EPTA:2023xxk} and references therein.
}  It is obtained as follows:
\begin{align}
    \Omega_\text{GW}^\text{(merger)}(\eta_0 , f) = \frac{f}{\rho_\text{c}} \int_0^{z_\text{sup}} \mathrm{d} z \, \frac{R(z)}{(1 + z) H(z)} \frac{\mathrm{d}E(f_\text{s})}{\mathrm{d}f_\text{s}} , \label{Omega_GW_merger}
\end{align}
where $\rho_\text{c}$ is the critical density, $z_\text{sup}$ is the upper bound of the redshift integral, $R$ is the merger rate of PBH binaries per comoving volume, $f_\text{s} = (1 + z) f$ is the frequency in the source frame, and $\mathrm{d}E/\mathrm{d}f_\text{s}$ is the GW spectrum from a merger in the source frame~\cite{Ajith:2007kx, Ajith:2009bn}. The expressions of the upper bound $z_\text{sup}$, the merger rate $R$, and the energy spectrum $\mathrm{d}E/\mathrm{d}f_\text{s}$ are given in Appendix~\ref{sec:merger_formulas}. 
For the calculation of the merger rate, we follow Ref.~\cite{Sasaki:2018dmp} and neglect the effects of PBH cloud, which may disrupt the binary, once the binary is formed. See  Ref.~\cite{Raidal:2018bbj} for the effects of the surrounding PBHs, which may significantly affect the merger rate.

\begin{figure}[tbh]
    \centering
    \includegraphics[width=0.6\textwidth]{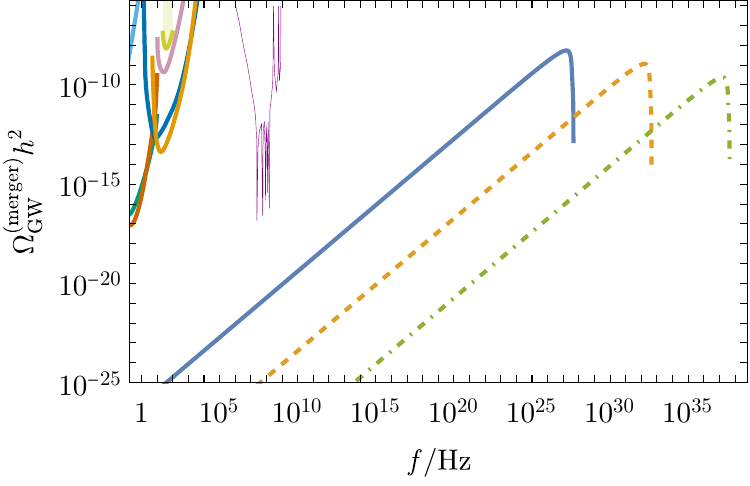}
    \caption{
    The spectra of the GWs from mergers of memory-burdened PBH binaries. The PBHs are assumed to comprise the totality of dark matter.  From left to right, $M_\text{PBH,ini}/\mathrm{g} = 10^{10}$ (the blue solid line), $10^5$ (the orange dashed line), and 1 (the green dot-dashed line) with $M_\text{PBH} = M_\text{PBH,ini} /2$. The monochromatic mass function is assumed for simplicity. The same sensitivity curves as in Fig.~\ref{fig:GW_spectra} are also shown. 
 }
    \label{fig:GW_spectra_merger}
\end{figure}

The spectrum of $\Omega_\text{GW}^\text{(merger)}$ is shown in Fig.~\ref{fig:GW_spectra_merger}. 
The spectral shape is the same as the previous work in the literature. It follows the power-law $\Omega_\text{GW}^\text{(merger)} \propto f^{2/3}$~\cite{Phinney:2001di, Ajith:2007kx, Ajith:2009bn} below the peak frequency. 
The peak frequency is $f_\text{peak} = 2 \times 10^{27} \, (M_\text{PBH,ini}/(10^{10}\, \mathrm{g}))^{-1}\, \mathrm{Hz}$, and the intensity at the peak is $\Omega_\text{GW}^\text{(merger)}(f_\text{peak})h^2 = 6 \times 10^{-9} 
 (M_\text{PBH,ini}/(10^{10}\,\mathrm{g}))^{5/37}$.
Currently, there are no known methods to detect such extremely high-frequency GWs. For recent developments toward the detection of high-frequency GWs, see Refs.~\cite{Dolgov:2012be, Domcke:2020yzq, Ejlli:2019bqj, Aggarwal:2020olq, Herman:2020wao, Berlin:2021txa,  Domcke:2022rgu, Franciolini:2022htd, Tobar:2022pie, Ahn:2023mrg, Ito:2019wcb, Ito:2020wxi, Ito:2022rxn, Ito:2023fcr,Liu:2023mll,Ito:2023nkq}. 

%====================================================================
%====================================================================

\section{Conclusions\label{sec:conclusion}}

Despite active experimental programs, we do not have clear evidence of dark matter production at colliders or dark matter detection in direct and indirect detection experiments so far. If this situation continues, the nightmare scenario of dark matter, namely the scenario with only gravitationally interacting dark matter, will become important. Macroscopic dark matter is special in this context since they can offer some characteristic observational handles as in the case of the PBH scenario.  One can imagine a pessimistic situation in which the conventional PBH dark matter mass window is observationally excluded. In this case, the PBHs beyond the semiclassical regime, namely the memory-burdened PBHs, become observationally more relevant in addition to their intrinsic importance. 
In this context, the induced GW signal associated with the memory-burdened PBH dark matter gives us important information: the mass of the PBHs. It can also be used for a consistency check of the memory-burdened PBH dark matter scenario.  

We have discussed possibilities to observationally confirm, in principle, the memory burden effect working on PBH dark matter.  In particular, the merger-originated GW signal is expected although the detailed form depends on the nonlinear dynamics of the memory-burdened PBH dark matter.  The peak frequency is $f_\text{peak} = 2 \times 10^{27}\, (M_\text{PBH, ini}/(10^{10}\, \mathrm{g}))^{-1}  \, \mathrm{Hz} $, which is far beyond the current limitation of the GW detection. 

In contrast, we can also exclude the memory burden effect if the induced GWs associated with the light PBHs turn out to be stronger than expected, up to the non-minimal scenario that instead constrains the generation mechanism of the baryon asymmetry of the Universe.  The observational constraint on the memory burden effect would be a valuable hint for black holes beyond the semi-classical regime and quantum gravity.

Taking into account the memory burden effect, we have studied the GWs induced by the enhanced primordial curvature perturbations that resulted in the formation of PBHs lighter than $10^{10}\,\mathrm{g}$.  From the requirement that the PBH abundance explains the dark matter abundance, the intensity of the induced GWs is fixed precisely whereas the frequency of the induced GWs corresponds to the initial mass of the PBHs. The peak frequency of the induced GWs is $f_\text{peak} = 1\times 10^{3} \, (M_\text{PBH}/(10^{10}\,\mathrm{g}))^{-1/2}\, \mathrm{Hz}$ and the peak intensity is $\Omega_\text{GW}(f_\text{peak})h^2 = 7 \times 10^{-9}$.  Although the frequency is high, there is the IR tail extending to lower frequencies.  We have found that the induced GWs can be tested by future GW observations such as Cosmic Explorer.  In conclusion, GWs give us clues about black holes beyond the semiclassical regime as well as about dark matter in the Universe.

\section*{Note Added}
Soon after our paper, Ref.~\cite{Jiang:2024aju} appeared on arXiv, which also discusses the induced GWs to constrain the PBH abundance in the context of the memory burden effect.  There are some differences about the considered power spectra of curvature perturbations and the prescription to calculate the PBH abundance.  While we focus more on the future prospects, they focus more on the current constraints from LIGO-Virgo O1-O3 runs.  In this work, we additionally studied ways to either confirm or exclude the memory burden effect in Sec.~\ref{sec:tests}. 

\section*{Acknowledgment}
K.K.~and T.T.~thank Sai Wang for discussions on the GWs from mergers of PBHs. This work was in part supported by MEXT KAKENHI Grants No.~JP23KF0289, No.~JP24H01825, No.~JP24K07027 (K.K.), and No.~24H02244 (T.T.Y.). T.T.Y.~was supported also by the Natural Science Foundation of China (NSFC) under Grant No.~12175134 as well as by World Premier International Research Center Initiative (WPI Initiative), MEXT, Japan. T.T.'s work was supported by the 34th Academic research grant FY 2024 (Natural Science) No.~9284 from DAIKO FOUNDATION and by the RIKEN TRIP initiative (RIKEN Quantum).

\appendix 

\section{Gravitational waves from mergers of PBH binaries}\label{sec:merger_formulas}
In this appendix, we summarize the formulas to calculate the spectrum of the GWs from mergers of PBH binaries.  We follow Ref.~\cite{Sasaki:2018dmp} as well as Appendixes B and C of Ref.~\cite{Wang:2019kaf}. See also Appendix B of Ref.~\cite{Inomata:2020lmk}. For simplicity, we assume a monochromatic PBH mass spectrum and the homogeneous spatial distribution of the PBHs. Also, we use the late-time PBH mass (in the memory burden regime) and do not take into account the time dependence of the PBH mass.  This can be justified since light PBHs quickly enter the memory burden regime and the evaporation in the memory burden regime is negligible for sufficiently large values of $n_\text{MB}$.  

For the convenience of the reader, we show Eq.~\eqref{Omega_GW_merger} again:
\begin{align*}
    \Omega_\text{GW}^\text{(merger)}(\eta_0 , f) = \frac{f}{\rho_\text{c}} \int_0^{z_\text{sup}} \mathrm{d} z \, \frac{R(z)}{(1 + z) H(z)} \frac{\mathrm{d}E(f_\text{s})}{\mathrm{d}f_\text{s}}.
\end{align*}
The upper boundary of the integral is $z_\text{sup} = \min [ z_\text{max}, (f_3 / f) - 1]$ where $z_\text{max}$ is the redshift at formation of PBHs, and $f_3$ is shortly given below. Practically, we can set $z_\text{sup} = (f_3/f) - 1$. 

Assuming the binary formation due to the torque from the nearest (third) PBH, the merger rate per comoving volume is~\cite{Sasaki:2018dmp}
\begin{align}
    R(z) = & \frac{f_\text{PBH} \Omega_\text{CDM} \rho_\text{c}}{M} \frac{\mathrm{d} P_t}{\mathrm{d} t},
\end{align}
where the differential merger rate at time $t$ for a PBH binary is~\cite{Sasaki:2018dmp}
\begin{align}
   \frac{\mathrm{d} P_t}{\mathrm{d} t} =& \frac{3}{58 t} \times \begin{cases}
       \left(\frac{t}{T} \right)^{3/37} - \left( \frac{t}{T} \right)^{3/8} & (t < t_\text{c}) \\
       \left(\frac{t}{T}\right)^{3/8} \left( \left(\frac{t}{t_\text{c}}\right)^{-29/56}\left( \frac{4\pi f_\text{PBH}}{3} \right)^{-29/8} - 1\right) & (t \geq t_\text{c})
   \end{cases}, 
\end{align}
where $T = \frac{729}{340 \pi^2 (1 + z_\text{eq})^4 (4 \pi f_\text{PBH}^{16} M_\text{PBH}^5 \rho_\text{c}^4 / 3)^{1/3}}$ and $t_\text{c} = (4\pi f_\text{PBH}/3)^{37/3} T$.

The energy spectrum of the GWs at emission is~\cite{Ajith:2007kx, Ajith:2009bn}
\begin{align}
    \frac{\mathrm{d}E}{\mathrm{d}f_\text{s}}(f_\text{s}) =& \frac{M_\text{c}^{5/3}}{12} \times \begin{cases}
        f_\text{s}^{-1/3} & (f_\text{s} \leq f_1 ) \\
        f_1^{-1} f_\text{s}^{2/3} & (f_1 < f_\text{s} \leq f_2) \\
        \frac{f_1^{-1}f_2^{-4/3}  f_\text{s}^2}{\left( 1 + \frac{4(f_\text{s} - f_2)^2}{\sigma^2} \right)^2} & (f_2 < f_\text{s} \leq f_3) \\
        0 & (f_3 < f_\text{s}) 
    \end{cases},
\end{align}
where $M_\text{c} = 2^{-1/3} M_\text{PBH}$ is the chirp mass, $f_1 = 0.4499 M_\text{PBH}^{-1}$, $f_2 = 1.026 M_\text{PBH}^{-1}$, $f_3 = 1.401 M_\text{PBH}^{-1}$ and $\sigma = 0.2381 M_\text{PBH}^{-1}$, where we have simplified the expressions assuming the monochromatic spectrum of the PBHs. Remember that we use $8\pi G = 1$ rather than $G= 1$.  

As mentioned briefly in the main text, the comoving merger rate is modified by surrounding PBHs because a binary may be disrupted by the surrounding PBHs. This issue was studied in Ref.~\cite{Raidal:2018bbj} with $N$-body simulations. Unfortunately, the analytic estimate in Ref.~\cite{Raidal:2018bbj} is not directly applicable for the case $f_\text{PBH} = 1$, but the authors of Ref.~\cite{Raidal:2018bbj} provided a conservative estimate of the merger rate for $f_\text{PBH}=1$ and $M_\text{PBH} \approx 30 M_\odot$.  The merger rate derived as above and the conservative estimate in Ref.~\cite{Raidal:2018bbj} deviate more for smaller PBH mass, which may suggest that the latter may be too conservative after the extrapolation of the PBH mass over more than 20 orders of magnitude into our parameter space $M_\text{PBH} \leq 10^{10}\, \mathrm{g}$.  If such an extrapolation is valid, the merger rate is substantially suppressed, so are the resulting GWs.  

Precisely speaking, the GW spectrum should follow the universal IR spectrum ($\Omega_\text{GW}\propto f^3$) well below the frequency corresponding to the inspiral frequency just after the binary formation~\cite{Cai:2019cdl}. Using the estimate of the IR cutoff frequency scale in Ref.~\cite{Cai:2019cdl}, one can see that Fig.~\ref{fig:GW_spectra_merger} is not affected in its plot region. Therefore, we neglect this effect for simplicity. 
For more on this point, see the discussions in Refs.~\cite{Cai:2019cdl, Inomata:2020lmk}. The spectrum of the merger-originated GWs is far below the observational sensitivity around and below the frequency range to which ground-based interferometers are sensitive.

%----------------------------------------------------------------------
\section{SNR contours for the other observations}\label{sec:SNR}
\begin{figure}[tbhp]
    \centering
    \subcaptionbox{ET}{\includegraphics[width=0.49\textwidth]{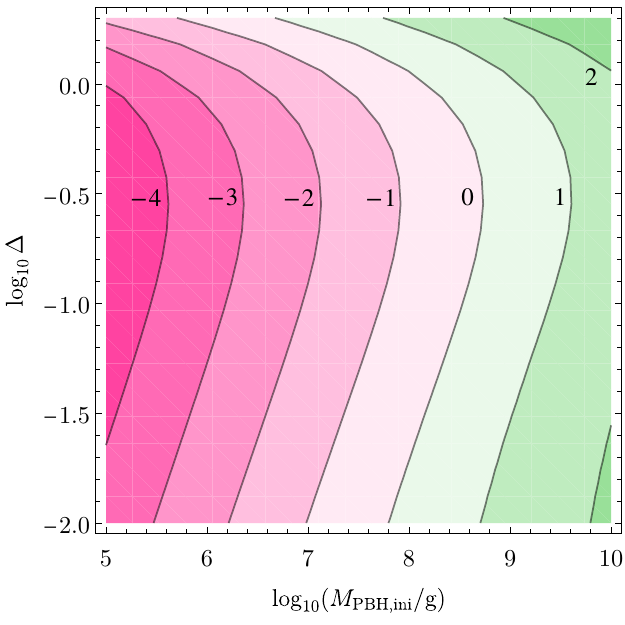}}
    \subcaptionbox{BBO}{\includegraphics[width=0.49\textwidth]{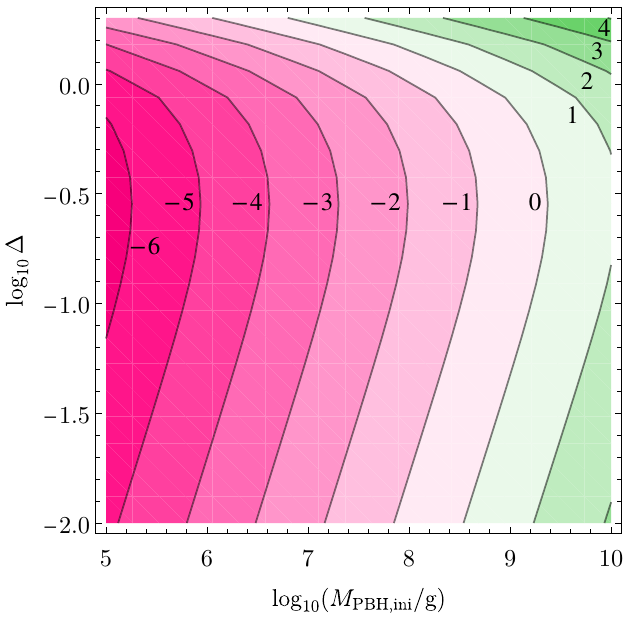}} \\
    \vspace{4mm}
    \subcaptionbox{DECIGO}{\includegraphics[width=0.49\textwidth]{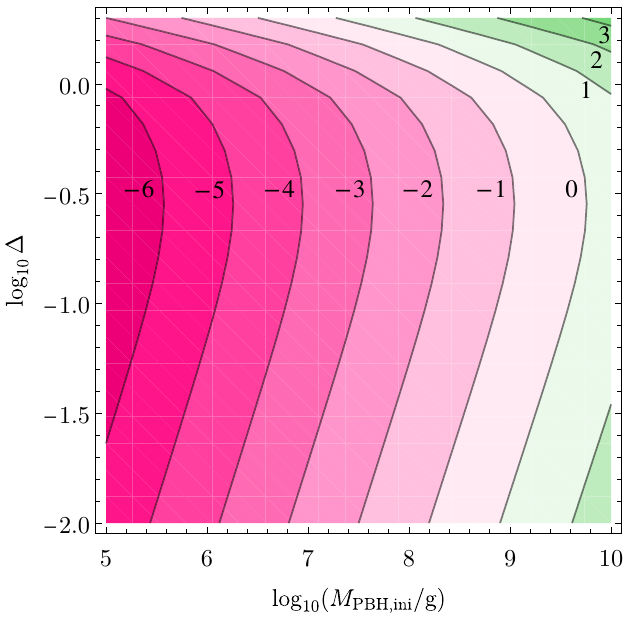}}
    \subcaptionbox{HLVK}{\includegraphics[width=0.49\textwidth]{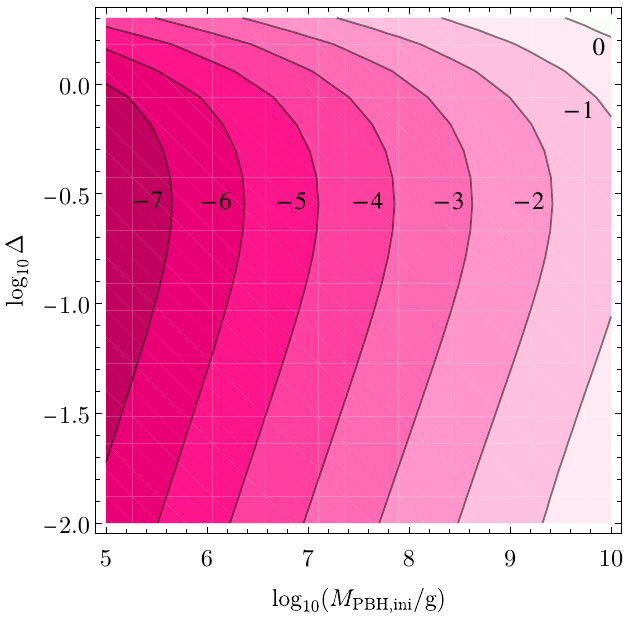}}
    \caption{
    The contour plots of $\log_{10} \text{SNR}$ for 1-year observations of (a) ET, (b) BBO, (c) DECIGO, and (d) HLVK on the plane of $\log_{10}(M_\text{PBH,ini}/\text{g})$ and $\log_{10}\Delta$. The border between the pinkish (low SNR) and greenish (high SNR) regions corresponds to $\text{SNR} = 1$ and the $\log_{10}\text{SNR}$ changes by $1$ for the adjacent contours. 
    The corresponding figure for CE is shown on the right panel of Fig.~\ref{fig:SNR}
 }
    \label{fig:SNR_complete}
\end{figure}

In this appendix, we show the contour plots of the $\log_{10}\text{SNR}$ for 1-year observations of the induced GWs for future GW observations: ET, BBO, DECIGO, and HLVK. See Fig.~\ref{fig:SNR_complete}. The case of CE is shown on the right panel of Fig.~\ref{fig:SNR}. 

%====================================================================
%====================================================================

\bibliographystyle{utphys}
\bibliography{memory_burden_GW}

\end{document}